\documentclass{ws-ijmpc}
\usepackage{color}

\begin{document}

\markboth{Hwee Kuan Lee and Yutaka Okabe}
{Multispin Coding Technique for Nonequilibrium Reweighting}

\catchline{}{}{}{}{}

\title{MULTISPIN CODING TECHNIQUE FOR NONEQUILIBRIUM REWEIGHTING
}

\author{HWEE KUAN LEE\footnote{Present address: 
Data Storage Institute, DSI Building, 
5 Engineering Drive 1, Singapore 117608} \ and YUTAKA OKABE}

\address{Department of Physics, Tokyo Metropolitan University, 
Hachioji, Tokyo 192-0397, Japan
}

\maketitle

\begin{history}
\received{Day Month Year}
\revised{Day Month Year}
\end{history}

\begin{abstract}
We present the multispin coding for the nonequlibrium reweighting method of
the Monte Carlo simulation, that was developed by the present authors.  As
an illustration, we treat the driven diffusive lattice gas model.  We use 
the multispin coding  technique both for the spin update and for the 
calculation of the histogram of incremental weights, which is needed in the
calculation of nonequlibrium reweighting.  All the operations are executed 
by the bitwise logical commands.  

\keywords{nonequilibrium; reweighting; Monte Carlo; 
lattice gas.}
\end{abstract}

\section{Introduction}

Monte Carlo simulations are used as standard techniques to investigate
statistical mechanical properties of many-body systems. The use of 
efficient algorithms is important, and much effort has been devoted to the
development of new Monte Carlo algorithms, such as cluster algorithms 
\cite{sw87,wolff89,hklee01,pcc} and extended ensemble methods 
\cite{berg91,Lee93,oliveira98,wl01,yko02}.

For systems with discrete symmetry, a refined treatment is possible for the
coding of the computer program. For instance, in the simulation of the 
Ising model, only one bit is required for storing the information of a 
single spin; a computer word can store the information of several spins at
the same time. The multispin coding technique developed by Bahnot 
{\it et al.} \cite{bhanot} and Michael \cite{michael} is based on this 
fact, and the Ising spins of 64 systems are put in a single computer word 
when using the 64-bit machine. The Monte Carlo spin flip process 
is executed with the logical commands, 
and 64 Ising systems are updated simultaneously with
a single random-number sequence. Accordingly, the computation time is 
reduced remarkably. For 64 systems, one may assign either systems having 
different parameters, for example, the temperature, the external field, 
etc., or systems having the same parameters but with different random 
number sequences.
The multispin coding has been successfully used for the study of the Ising
model on the regular lattices \cite{kikuchi87,ito88} and that on the 
quasicrystal \cite{okabe88}. The three-state random Potts model together 
with the block-spin transformation was also formulated using the multispin
coding \cite{kikuchi95}. The multispin coding was applied not only to the 
single spin flip dynamics but also to the Kawasaki spin exchange dynamics
\cite{gawlinski,roland,zhang}.

The histogram reweighting \cite{ferrenberg88} is a powerful method to 
investigate the equilibrium properties with simulations; only simulation at
a single temperature is required to obtain information for a range of 
temperatures. Recently, the present authors~\cite{hklee05a} have extended the
idea of reweighting to the case of nonequilibrium systems based on the 
Sequential Importance Sampling \cite{doucet,liubook}.
A sequence of the micro-states, or a path, is generated, and the average 
over a single path is performed to calculate thermodynamic quantities 
for a standard Monte Carlo simulation. 
In nonequilibrium reweighting, on the contrary, many paths are first
sampled with a trial distribution, and thermodynamic quantities are 
calculated based on the relative probability between the trial distribution
and the target distribution. The relative probability is called ``weights"
in literature, which we shall use hereafter. The nonequilibrium reweighting
technique was used for the nonequilibrium relaxation of the Ising model, 
and the dynamical properties of the phase transition are discussed 
\cite{hklee05a}. Moreover, the present authors \cite{hklee05b} applied the 
nonequilibrium reweighting to the study of the nonequilibrium steady states
\cite{zia,dickman}; the driven diffusive lattice gas model proposed by
Katz, Lebowitz and Spohn (KLS) \cite{katz} was employed as an example of 
the system showing the phase transition to the nonequilibrium steady state.

In this paper, we present the multispin coding of the nonequlibrium 
reweighting. It greatly saves computation time. We pick up the driven 
diffusive lattice gas model, the KLS model \cite{katz}, to illustrate the 
multispin coding, but the formulation is general.  Of course, it can be 
used for the Ising model, which is simpler.

We organize the rest of the paper as follows.  In Sec. \ref{sec:model}, we
briefly explain the KLS model.  In Sec. \ref{sec:reweighting}, 
a short description of the nonequilibrium reweighting method is given. 
Sec. \ref{sec:multispin}, which is the main part of this paper, gives the 
illustration of the multispin coding.  In Sec. \ref{sec:results}, we give
the result of the calculation. The final section is devoted to summary and
discussions.

\section{Model}
\label{sec:model}

We use the KLS model~\cite{katz} for an illustration of the nonequilibrium
reweighting method. This model consists of a {\it half filled} lattice gas 
on a periodic $L_x\times L_y$ square lattice. Its Hamiltonian is given by
%
%
\begin{equation}
{\cal H} = -4 \sum_{\langle i j ,i' j' \rangle} n_{ij} n_{i'j'}
\end{equation}
%
where the sum of lattice sites $\{ i j\}$ is over nearest neighbors, and the
variable at the lattice site $n_{ij} = 1$ if the site is filled and 
$n_{ij}=0$ if the site is empty. For the Monte Carlo update, particles are
allowed to hop to an empty nearest neighbor site with the Metropolis rate,
%
\begin{equation}
T_{\beta,E}(\sigma' | \sigma) = 
   \min[ 1, \exp(-\beta (\Delta {\cal H} - \epsilon E)) ]
\label{eq:trate}
\end{equation}
%
where $\sigma$ and $\sigma'$ are the system configurations before and after
the hop, $\Delta \cal H$ represents the change in energy due to the hop,
$E$ is a constant driving force, $\epsilon = -1, 0$ or $+1$ depending on 
whether the hop is against, orthogonal or along the direction of the 
drive, and $\beta = 1/T$ is the inverse temperature of the heat bath. The 
$L_y$ direction  is taken as the direction of the drive.

The KLS model exhibits an order-disorder second order phase transition. The
ordered phase consists of strips of high- and low-density domains in the 
direction of the drive. In the final steady state, a single strip of high 
density domain is formed~\cite{hurtado}. The density profile along the
direction of the drive can be regarded as the order parameter, and it can 
be defined as
%
\begin{equation}
\rho = \frac{1}{(L_x/2)} \sum_{j=1}^{L_x} \left| \frac{1}{L_y} 
\sum_{i=1}^{L_y} n_{ij} - \frac{1}{2} \right|
\label{order}
\end{equation}

\section{Nonequilibrium Reweighting}
\label{sec:reweighting}

In a Monte Carlo simulation, the system configuration changes with time. In
each Monte Carlo step, a randomly chosen particle is attempted to hop to a 
nearest neighbor site and the attempt is either accepted or rejected. Let 
$\sigma_1$ be the initial system configuration and let $\sigma_2$ be the 
system configuration of the second Monte Carlo step, and $\sigma_3$ be the
system configuration of the third Monte Carlo step and so on. In this way,
a Monte Carlo simulation can be represented by a series of system 
configurations and we define a ``path" of the simulation as follows,
%
%
\begin{equation}
\vec{x}_t = ( \sigma_1, \cdots, \sigma_{t-1}, \sigma_t)
\end{equation}
%
%
Many paths are sampled and the thermal average of a quantity $Q$ may be
calculated by averaging over these paths, $\langle Q(t) \rangle_{\beta,E} 
= (1/n) \sum_{j=1}^n Q(\vec{x}^j_t)$,
where the sum is over all paths indexed by $j$, $\beta$ and $E$ are the
inverse temperature and drive used in the Monte Carlo sampling process. The
objective of reweighting is to calculate the thermal averages of $Q$ at
another temperature $\beta'$ and drive $E'$. This can be achieved by using
appropriate weights $w$,
%
\begin{equation}
\langle Q(t) \rangle_{\beta',E'} = \sum_{j=1}^n w_t^j Q(\vec{x}_t^j) /
\sum_{j=1}^n w_t^j
\end{equation}
%
To calculate the weights, the following steps are implemented:
%
%
\begin{enumerate}
\item Assume that a path $\vec{x}_t^j$ up to some time $t$ is sampled from
      a simulation at $\beta$ and $E$.
\item In the next Monte Carlo step, choose a pair of neighboring sites at
      random. If one of the two sites is empty, perform a Kawasaki exchange
      between the two sites with the rate $T_{\beta,E}({\sigma'}^j | 
      \sigma_t^j)$. ${\sigma'}^j $ represents the trial system configuration 
      after the move.
\item The trial move is either accepted or rejected. In each case, we 
      define an incremental weight $\delta w^j$,
  \begin{itemize}
  \item[a)] If the move is accepted, 
        $\sigma_{t+1}^j = {\sigma'}^j $ and
        $\delta w^j = T_{\beta',E'}({\sigma'}^j |\sigma_t^j)/ 
                      T_{\beta ,E }({\sigma'}^j |\sigma_t^j)$.
  \item[b)] If the move is rejected, 
        $\sigma_{t+1}^j = \sigma_t^j$ and
        $\delta w^j = [ 1- T_{\beta',E'}({\sigma'}^j |\sigma_t^j) ]/ 
                      [ 1 - T_{\beta ,E }({\sigma'}^j |\sigma_t^j)] $.
  \end{itemize}
\item The weights at $t+1$ are given by multiplication: 
      \begin{equation}
      w_{t+1}^j = w_{t}^j \times \delta w^j
      \end{equation}
%
\end{enumerate}
For each path $j\in \{ 1, \cdots, n \}$ these steps are repeated until a 
predetermined maximum Monte Carlo time is reached.

\section{Coding Techniques}
\label{sec:multispin}

We shall illustrate multispin coding for reweighting with infinite drive;
generalization to finite drive is straightforward. We use bitwise 
operations to update our system configurations and to calculate the 
weights. On a 64-bit machine, 64 systems are simulated in one run. 
For an $L_x \times L_y$ lattice, we use $L_x L_y$ words to represent 
simultaneously $64$ lattices with each bit in the word representing one 
lattice site. 
Our multispin implementation is 
done systematically using truth tables. Firstly, we construct the 
appropriate truth tables for various tasks, and then we implement them 
using bitwise operations in the computer.
%

\begin{table}
\tbl{Mapping of $\Delta {\cal H}-\epsilon E$ to 
a four-bit pattern $\vec M$. The transition rates (Eq.~(\ref{eq:trate})) are
also given.}
{\begin{tabular}{@{}ccc@{}} \toprule
$\Delta {\cal H} - \epsilon E$  & 
Transition rate & 
$\vec{M} = (M_3,M_2,M_1,M_0)$  \\ \colrule
$\infty$  &  $0$               & 0 0 0 0  \\ 
$ 12   $  &  $\exp(-12 \beta)$ & 0 0 0 1  \\ 
$  8   $  &  $\exp(-8 \beta)$  & 0 0 1 0  \\ 
$  4   $  &  $\exp(-4 \beta)$  & 0 0 1 1  \\ 
$  0   $  &  1                 & 0 1 0 0  \\ 
$ -4   $  &  1                 & 0 1 0 1  \\ 
$ -8   $  &  1                 & 0 1 1 0  \\ 
$ -12  $  &  1                 & 0 1 1 1  \\ 
$-\infty$ &  1                 & 1 0 0 0  \\ 
\botrule
\end{tabular}
\label{tbl:trate}
}
\end{table}
%

\subsection{Multispin coding  of the Kawasaki exchange}

\begin{figure}
\centerline{\includegraphics[width=6.4cm]{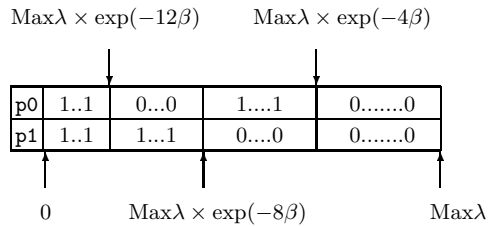}}
\caption{The entries of {\tt p0} and {\tt p1} are assigned 
according to the transition rates.}
\label{fig:jtbl}
\end{figure}

In the Kawasaki exchange, a pair of neighbor sites is chosen and if one
site is empty and the other site is filled, an attempt is made for the 
particle in the filled site to hop to the empty site with a rate given by 
Eq.~(\ref{eq:trate}). In the coding implementation, we should note that 
possible values of $\Delta {\cal H} - \epsilon E$ for infinite drive 
are $\infty, 12, 8, 4, 0, -4, -8, -12$, and $-\infty$. 
For example, when the attempt is made to move a 
particle against the drive, $\Delta {\cal H} - \epsilon E = \infty$ and if
the move is along the drive, $\Delta {\cal H} - \epsilon E = -\infty$. If 
the move is orthogonal to the drive, $\epsilon=0$ and $\Delta {\cal H} = 
-12,-8,-4,0,4,8$ or $12$ depending on the local particle configurations 
of nearest neighbors to the chosen sites. We map the values of $\Delta 
{\cal H} - \epsilon E$ into a four-bit pattern 
$\vec{M} = (M_3,M_2,M_1,M_0)$, which is illustrated in Table 
\ref{tbl:trate}; there, we also give the transition rate 
(Eq.~(\ref{eq:trate})) for each $\Delta {\cal H} - \epsilon E$. 
In the multispin coding, 
we introduce a dummy variable $p$, which takes $0$ to $3$ in the 
present case, to implement the acceptance and rejection procedure. Two 
arrays \verb+p0+ and \verb+p1+ 
are allocated for representing the probability of appearance of $p$. 
The array variable $\lambda$ for \verb+p0+[$\lambda$] and 
\verb+p1+[$\lambda$] takes 0 to Max$\lambda-1$.
For the size of array, Max$\lambda$, we choose a large enough number, for
example $2^{24}$. The values of each bit in \verb+p0+ and \verb+p1+ are set
according to the transition rates as shown in Fig. \ref{fig:jtbl}.
Bits in the $\lambda$th element of the arrays, when put together as
(\verb+p1+[$\lambda$], \verb+p0+[$\lambda$]), form a binary representation
of the numbers $3(11), 2(10), 1(01)$ and $0(00)$. The arrays 
\verb+p0+ and \verb+p1+ are related to the transition 
probability distribution as follows.  
Let $\lambda$ be a random number between $0$ and Max$\lambda -1$, 
then the probability of getting the array elements with the bit pattern 
(\verb+p1+[$\lambda$], \verb+p0+[$\lambda$]) $\ge 2$ is $\exp(-8\beta)$, 
for example. 
We build the arrays \verb+p0+ and \verb+p1+ bit-by-bit and shuffle each bit 
independently. Note that the sequence of shuffling for each $n$th bit in 
\verb+p0+ and \verb+p1+ {\it must} be the same for both arrays.

%
%
%

Then, the actual Monte Carlo procedure in the multispin coding 
can be summarized as follows;
%
%
\begin{enumerate}
\item Pick a pair of neighboring sites $n_{ij}$, $n_{i'j'}$ and 
      perform a Kawasaki exchange trial.
\item From $\Delta {\cal H} - \epsilon E$, determine $\vec{M} = 
      (M_3,M_2,M_1,M_0)$.
\item Let $\lambda$ be a random number generated uniformly between $0$ and
      Max$\lambda -1$.
\item Assign an acceptance bit $A=1$ (acceptance) if
      \begin{equation} 
      8 \times M_3 + 4 \times M_2 + 2\times M_1 + 
      M_0 + 2\times {\verb+p1+}[\lambda] + {\verb+p0+}[\lambda] \ge 4.
      \end{equation}
      Otherwise assign $A=0$ (rejection). Assignment of $A$ can be 
      implemented using Table \ref{tbl:accept}. 
\item Finally, the Kawasaki particle exchange is performed using 
      $XOR (\oplus)$ operations of $A$ with both neighboring lattice sites.
      %
      \begin{equation}
      n_{ij} \leftarrow n_{ij} \oplus A \mbox{\hspace{3mm}and\hspace{3mm}}
      n_{i'j'} \leftarrow n_{i'j'} \oplus A 
      \end{equation}
      %
\end{enumerate}
%
%
\begin{table}
\tbl{Truth table for acceptance decision. $A$ is a variable defined 
such that $A=1$ represents acceptance and $A=0$ represents rejection.}
{\begin{tabular}{@{}cccc|cccc|cccc|cccc@{}} \toprule
$\vec{M}$  &  {\tt p1}  &  {\tt p0}  &  $A$  & 
$\vec{M}$  &  {\tt p1}  &  {\tt p0}  &  $A$  & 
$\vec{M}$  &  {\tt p1}  &  {\tt p0}  &  $A$  & 
$\vec{M}$  &  {\tt p1}  &  {\tt p0}  &  $A$  
\\ \colrule
0000 & 0 & 0 & 0 & 0000 & 0 & 1 & 0 & 0000 & 1 & 0 & 0 & 0000 & 1 & 1 & 0  \\ 
0001 & 0 & 0 & 0 & 0001 & 0 & 1 & 0 & 0001 & 1 & 0 & 0 & 0001 & 1 & 1 & 1  \\ 
0010 & 0 & 0 & 0 & 0010 & 0 & 1 & 0 & 0010 & 1 & 0 & 1 & 0010 & 1 & 1 & 1  \\ 
0011 & 0 & 0 & 0 & 0011 & 0 & 1 & 1 & 0011 & 1 & 0 & 1 & 0011 & 1 & 1 & 1  \\ 
0100 & 0 & 0 & 1 & 0100 & 0 & 1 & 1 & 0100 & 1 & 0 & 1 & 0100 & 1 & 1 & 1  \\ 
0101 & 0 & 0 & 1 & 0101 & 0 & 1 & 1 & 0101 & 1 & 0 & 1 & 0101 & 1 & 1 & 1  \\ 
0110 & 0 & 0 & 1 & 0110 & 0 & 1 & 1 & 0110 & 1 & 0 & 1 & 0110 & 1 & 1 & 1  \\ 
0111 & 0 & 0 & 1 & 0111 & 0 & 1 & 1 & 0111 & 1 & 0 & 1 & 0111 & 1 & 1 & 1  \\ 
1000 & 0 & 0 & 1 & 1000 & 0 & 1 & 1 & 1000 & 1 & 0 & 1 & 1000 & 1 & 1 & 1  \\ 
\botrule
\end{tabular}
\label{tbl:accept}
}
\end{table}

\subsection{Multispin coding of nonequilibrium reweighting}

Implementation of multispin coding for reweighting is much easier than that
of the Kawasaki particle exchange. Possible values of incremental weights 
$\delta {w_i}$ are
%
\begin{equation}
\begin{array}{rcl}
\delta {w_0}&=&1        \\
\delta {w_1}&=&\exp(-12 (\beta'-\beta))  \\
\delta {w_2}&=&\exp(-8 (\beta'-\beta))   \\
\delta {w_3}&=&\exp(-4 (\beta'-\beta))   \\
\delta {w_4}&=& ( 1 - \exp(-12 \beta')) / ( 1 - \exp(-12 \beta) )  \\
\delta {w_5}&=& ( 1 - \exp(-8 \beta') ) / ( 1 - \exp(-8 \beta) )  \\
\delta {w_6}&=& ( 1 - \exp(-4 \beta') ) / ( 1 - \exp(-4 \beta) )
\end{array}
\end{equation}
%
The weights can then be written as a product of incremental weights,
%
\begin{equation}
w^j_t = (\delta {w_1})^{h_1^j(t)} (\delta {w_2})^{h_2^j(t)}
        (\delta {w_3})^{h_3^j(t)} (\delta {w_4})^{h_4^j(t)}
        (\delta {w_5})^{h_5^j(t)} (\delta {w_6})^{h_6^j(t)}
\label{eq:msw}
\end{equation}
%
where $h^j_1(t) \cdots h^j_6(t) $ are the number of hits on the incremental
weights $ \delta w_1 \cdots \delta w_6$ during the course of simulation 
from time $1$ to $t$. Note that $\delta w_0$ is irrelevant in 
Eq.~(\ref{eq:msw}). Hence calculation of weights has been reduced to 
accumulating histograms. In our implementation, we store the hits in a 
large array $dh_i(k) = 0 \mbox{ or } 1$ for $i=1, \cdots, 6$ and 
$k=1, \cdots, t_0$ for an arbitrarily chosen $t_0$. Weights and histograms 
are updated only once every $t_0$ steps using Eq.~(\ref{eq:msw}), and
$h^j_i (t) = h^j_i (t-t_0) + \sum_{k=1}^{t_0} dh_i(k)$.
For calculating $dh_i(k)$, we use the four-bit pattern $\vec{M}$ and 
acceptance decision $A$ defined earlier. Then for each $k = 1, \cdots, t_0$,
assign $dh_i(k)$ according to the truth table given in Table 
\ref{table:dh}. To calculate the summation $\sum_{k=1}^{t_0} dh_i(k)$ in 
the multispin coding, we use the same routine to compute the total magnetization
or energy. The bit-counting routine, \verb+BITCNT+, by Ito and 
Kanada~\cite{ito88} can be used. The essential point is that we do not have
to calculate the weights by multiplication at each time, 
but we only need to calculate the histograms $h_i(t)$, 
which are obtained by integer operations.

%
\begin{table}
\tbl{Truth table for the mapping from $(\vec{M},A)$ to $dh_i$, 
$i=1, \cdots, 6$.}
{
\begin{tabular}{@{}cccccccc@{}} \toprule
$\vec{M}$ & $A$ & $dh_1$&$dh_2$&$dh_3$&$dh_4$&$dh_5$&$dh_6$\\ \colrule
0001  &	 1     &  1    & 0    & 0    & 0    & 0    & 0 \\
0010  &	 1     &  0    & 1    & 0    & 0    & 0    & 0 \\
0011  &	 1     &  0    & 0    & 1    & 0    & 0    & 0 \\
0001  &	 0     &  0    & 0    & 0    & 1    & 0    & 0 \\
0010  &	 0     &  0    & 0    & 0    & 0    & 1    & 0 \\
0011  &	 0     &  0    & 0    & 0    & 0    & 0    & 1  \\ 
\botrule
\end{tabular}
\label{table:dh}
}
\end{table}
%

\section{Result}
\label{sec:results}

\begin{figure}
\centerline{\includegraphics[width=7.6cm]{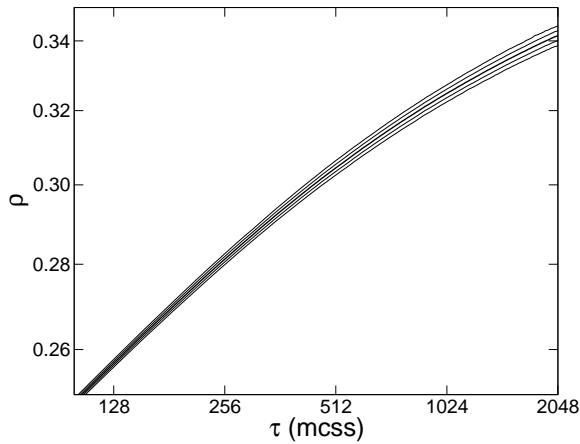}}
\caption{Order parameter with infinite drive on the $64\times 32$ lattice.
From top to bottom, values of $T$ are $3.150, 3.155, 3.160, 3.165$, and 3.170.}
\label{fig:infinite}
\end{figure}

We performed simulations with reweighting on the KLS model with infinite 
drive.   In Fig. \ref{fig:infinite}, as an example, 
we show the time variation of the
order parameter, Eq.~(\ref{order}), for $64\times 32$ lattice 
with infinite drive. Simulations 
were performed at $T=3.160$ and data reweighted to nearby temperatures,
$T=3.150, 3.155, 3.165$, and 3.170 (from top to bottom). 
Averages were taken over
$4.096\times 10^6$ samples.  We should note that the direct calculations 
at temperatures different from $T=3.160$ gave the consistent 
results as the reweighted ones within statistical errors, 
which shows the effectiveness of the reweighting. 
The detailed analysis of the phase transition based on the finite-size 
scaling was reported in a separate paper \cite{hklee05b}.

We here mention about the performance of our multispin coding. 
The computation time of the calculation with multispin coding 
for 64 systems is 40 to 60 times faster than 
that for independent 64 simulations 
without multispin coding for typical system size.  
This is a rough estimate because 
such a comparison depends on the optimization of program. 
The overhead of computation time to calculate the histogram $h_i(t)$ 
is the same order as the spin flip process, if we use the \verb+BITCNT+ 
routine~\cite{ito88} once per $N$ single spin flip steps. 
This is the same situation as the calculation of the thermal 
quantities. 
In this way, the calculation of weights by the histogram, Eq.~(\ref{eq:msw}), 
reduces the computation time.  
More important is that the accumulated errors will be much 
reduced for histogram calculations 
compared to the calculation with the multiplication of 
incremental weights each time. 

\section{Summary and Discussions}

We have formulated the nonequilibrium reweighting method that is convenient
for implementing multispin coding. In addition, we have shown how multispin
coding can be implemented for both particle hopping and nonequilibrium 
reweighting.  All the computations are performed in terms of integer 
with logical commands.  
As a result, a large increase of efficiency 
has been achieved.
We would like to remark that the nonequilibrium reweighting method
is general and may be applied to various models with different Monte Carlo
updates.  In this paper, we have treated the case of two component system 
with the Metropolis update.  Nonequilibrium reweighting can be 
applied to multi component systems, for example, $q$-state Potts models, 
and also to the heat-bath update.  A generalization of the formulation 
is necessary, which will be discussed elsewhere \cite{hklee05c}.

\section*{Acknowledgments}

The authors thank N. Ito for valuable discussions. 
This work is supported by a Grant-in-Aid for Scientific Research from the
Japan Society for the Promotion of Science. The computation of this work 
has been done using computer facilities of the Supercomputer Center, 
Institute of Solid State Physics, University of Tokyo.

\end{document}